\documentclass[12pt]{article} 

\usepackage{amssymb}
\usepackage{amsmath}
\usepackage{amscd}
\usepackage{latexsym}
\usepackage{graphicx}

\usepackage{cite}

\newcommand{\be}{\begin{equation}}
\newcommand{\ee}{\end{equation}}

\newcommand{\dlt}{\delta}
\newcommand{\prt}{\partial}
\newcommand{\br}{{\bf r}}
\newcommand{\bk}{{\bf k}}
\newcommand{\bS}{{\bf S}}
\newcommand{\ba}{{\bf a}}
\newcommand{\bM}{{\bf M}}

\newcommand{\bt}{\beta}
\newcommand{\vp}{\varphi}
\newcommand{\ep}{\varepsilon}
\newcommand{\al}{\alpha}
\newcommand{\ra}{\rightarrow}
\newcommand{\sgm}{\sigma}

\newcommand{\gm}{\gamma}
\newcommand{\om}{\omega}

\newcommand{\dgr}{\dagger}
\newcommand{\lbd}{\lambda}

\newcommand{\cD}{{\cal D}}

\newcommand{\cH}{{\cal H}}
\newcommand{\cF}{{\cal F}}
\newcommand{\cA}{{\cal A}}
\newcommand{\cB}{{\cal B}}
\newcommand{\rgl}{\rangle}
\newcommand{\lgl}{\langle}

\begin{document}

\begin{center}

{\Large{\bf Order indices and entanglement production in quantum systems} \\ [5mm]
Vyacheslav I. Yukalov$^{1,2}$} \\ [3mm]

{\it
$^{1}$Bogolubov Laboratory of Theoretical Physics, \\
Joint Institute for Nuclear Research, Dubna 141980, Russia \\
{\bf E-mail}: yukalov@theor.jinr.ru  \\ [3mm]

$^2$Instituto de Fisica de S\~ao Carlos, Universidade de S\~ao Paulo, \\
CP 369, S\~ao Carlos 13560-970, S\~ao Paulo, Brazil}

\end{center}

\vskip 2cm

\begin{abstract}
The review is devoted to two important quantities characterizing many-body 
systems, order indices and the measure of entanglement production. Order indices 
describe the type of order distinguishing statistical systems. Contrary to the 
order parameters characterizing systems in the thermodynamic limit and describing 
long-range order, the order indices are applicable to finite systems and classify 
all types of orders, including long-range, mid-range, and short-range orders. The 
measure of entanglement production quantifies the amount of entanglement produced 
in a many-partite system by a quantum operation. Despite that the notions of order 
indices and entanglement production seem to be quite different, there is an 
intimate relation between them, which is emphasized in the review.
\end{abstract}


\newpage

\section{Introduction}

Many-body systems can be described by different characteristics representing the 
main system properties. One of the pivotal notions is that of the type of order 
associated with the system state. The order is usually quantified by order parameters 
(see, e.g., \cite{Landau_1,Haar_2,Bogolubov_3}). The order parameters, as is well 
known, are rigorously defined in the thermodynamic limit, while for finite systems 
they, strictly speaking, degenerate to zero \cite{Bogolubov_3}. Here we mean a 
mathematically rigorous definition of order parameters, although in practical 
experiments and in numerical modeling one can still see the order parameters behave 
as if the thermodynamic limit is reached, when the system size is much larger than 
the correlation length. 

At the same time, large many-body systems, even being finite, can posses a kind of 
order that is not of a long-range type, but rather it is a quasi-long-range or 
algebraic order. In addition, some two-dimensional systems exhibit quasi-long-range order 
below the Berezinskii--Kosterlitz--Thouless transition \cite{Berezinskii_4,Kosterlitz_5}.    

Order indices for density matrices were introduced in Refs. 
\cite{Coleman_6,Coleman_7,Coleman_8,Coleman_9} as a general tool for quantifying all 
types of order, whether long-range, mid-range, or short-range. These characteristics 
can be defined for any statistical system, whether finite or in the thermodynamic 
limit. The notion of order indices can be generalized for arbitrary operators or 
matrices \cite{Yukalov_10}.    

Another important notion characterizing the state of a multi-partite system is 
entanglement widely employed in quantum information processing and quantum computing 
\cite{Williams_11,Nielsen_12,Vedral_13,Keyl_14,Horodecki_15,Guhne_16,Wilde_17}, as 
well as in the theory of quantum measurements and quantum decision theory 
\cite{Yukalov_S_87,Yukalov_18,Yukalov_19,Yukalov_20,Yukalov_S_88}. To be more precise, 
one has to distinguish between the entanglement of a state, or a statistical operator, 
or generally of an arbitrary operator, and the entanglement production describing the 
action of an operator on the given Hilbert space. The state entanglement characterizes 
the structure of a statistical operator, while the entanglement production by an 
operator describes the result of an operator action. These notions will be concretized 
below. 

The aim of the present review is two-fold. First, a survey of the notions of order 
indices and of entanglement production, and their applications for treating 
many-body systems, will be given. Second, we shall emphasize the interrelation between 
these characteristics. It turns out that the latter are intimately related with each 
other, so that the qualitative change of a state leads to the quantitative changes of 
order indices as well as of the entanglement production.  

Throughout the paper, the system of units is used where the Planck and Boltzmann 
constants are set to one.

\section{Order indices}

Below, we consider systems composed of $N$ parts, or particles, with $N$ being finite,
although it can be rather large. Suppose a collection $\{\hat{A}\}$ of trace-class 
operators acts on a Hilbert space $\mathcal{H}$, so that
\be
\label{1}
 0 < |\; {\rm Tr}_\cH \hat A \; | < \infty \; .
\ee

The order index of an operator $\hat{A}$ is defined as
\be
\label{2}
 \om(\hat A) \equiv 
\frac{\log||\; \hat A \; ||}{\log|\; {\rm Tr}_\cH \hat A\; | } \;  ,
\ee
where the base of the logarithm can be taken according to convenience, since the 
above definition does not depend on the choice of the base due to the property 
$\log_a(x) = \log_b(x) \log_a(b)$. In other words, the order index is the exponent 
connecting the norm and trace of an operator,
$$
||\; \hat A \; || = | \; {\rm Tr}_\cH \hat A\; |^{\om(\hat A)} \;  .
$$
In the case of a positive operator, 
$$
||\; \hat A \; || \leq  {\rm Tr}_\cH \hat A \qquad ( \hat A \geq 0 ) \; ,
$$        
because of which
\be
\label{3}
\om(\hat A) \leq 1 \qquad ( \hat A \geq 0 ) \;  .
\ee

As the norm, it is possible to take some of the Shatten norms
$$
 ||\; \hat A \; ||_p \equiv 
\left(  {\rm Tr}_\cH |\;\hat A \;|^p \right)^{1/p} \;  ,
$$
where
$$
 |\; \hat A \; | \equiv \sqrt{\hat A^+ \hat A} \qquad 
( p \in [ 1, \infty] ) \;  .
$$
Thus, for $p = 1$, we have the trace norm
$$
||\; \hat A \; ||_1 = {\rm Tr}_\cH |\;\hat A \;| \;   ,
$$
and for $p = 2$, we come to the Hilbert--Shmidt norm
$$
 ||\; \hat A \; ||_2 = 
\left(  {\rm Tr}_\cH |\;\hat A \;|^2 \right)^{1/2} \; .
$$
Below, it will be more convenient to accept the operator norm corresponding to 
$p=\infty$, which gives the operator norm
\be
\label{4}
||\; \hat A \; ||_\infty = \sup_\vp \; \frac{||\;\hat A\vp\;||}{||\;\vp\;||}
\qquad ( \vp \neq 0 ) 
\ee
generated by the vector norm
$$
||\;\hat A\vp\;|| = \sqrt{(\hat A\vp,\hat A\vp)} \; .
$$
Dealing with Hermitian operators, we get
\be
\label{5}
||\; \hat A \; ||_\infty = \sup_\vp \; \frac{(\vp,\hat A\vp)}{||\;\vp\;||}
\qquad ( \hat A^+ = \hat A ) \; .
\ee

Comparing two operators $\hat{A}_1$ and $\hat{A}_2$, we say that the operator 
having a larger order index is better ordered. In physical applications, the 
operator or matrix order indices characterize the type of order associated with 
the considered operators. Examples will be given below.

\section{Entangled structures}

One has to distinguish two different notions, entangled structures and entangling
operations. The first notion characterizes the property of such structures as wave 
functions, statistical operators (quantum states), and which can be generalized to 
arbitrary operators. The second notion describes the action of quantum operations 
on the members of a Hilbert space. To better explain the difference of these notions, 
in the present section we recall the main definitions concerning entangled structures, 
and in the next section we shall elucidate the meaning of the entanglement production 
by quantum operations.

\subsection{Entangled functions}
   
The notion of wave-function entanglement was introduced by Schr\"odinger 
\cite{Schrodinger_21,Schrodinger_22} with respect to quantum systems that can be 
separated into several subsystems. Then the system Hilbert space can be represented 
as the tensor product 
\be
\label{6}
\cH = \bigotimes_{i=1}^n \cH_i
\ee
of the subsystem Hilbert spaces. Wave functions of the whole system, pertaining to 
the space $\mathcal{H}$ can be separated into two classes, separable and entangled 
functions. Separable functions have the form of the product
\be
\label{7}
 \vp_{sep} = \bigotimes_i \vp_i \qquad ( \vp_i \in \cH_i ) \;  .
\ee
Entangled functions can be represented as the linear combinations
\be
\label{8}
\vp_{ent} = \sum_\al c_\al \bigotimes_i \vp_{i\al} \qquad 
( \vp_{i\al} \in \cH_i) \;   ,
\ee
where at least two coefficients $c_\alpha$ are not zero. From the physical point of 
view, separable wave functions are rather exceptional, being attributed to quantum 
subsystems that have never interacted and are distinguishable \cite{Schiff_23}. 
Generally, subsystems are characterized by entangled wave functions 
\cite{Yukalov_24,Yukalov_25}.   

The collection of all separable functions forms a disentangled set
\be
\label{9}
 \cD = \{ \vp_{sep} \in \cH \} \; .
\ee
All entangled functions compose an entangled set
\be
\label{10}
 \cH \setminus \cD = \{ \vp_{ent} \in \cH \} \;  .
\ee
Considering wave functions, one keeps in mind the functions normalized to one, but 
in principle, the same definitions are valid for non-normalized functions that are 
the members of the Hilbert space $\mathcal{H}$.

\subsection{Entangled states}

Generally, quantum states are represented by statistical operators. For pure states,  
statistical operators can be written as 
\be
\label{11}
 \hat\rho = |\; \vp\; \rgl \lgl \; \vp \; | \;  ,
\ee
where the standard bra-ket notation is used. Separable wave functions compose separable 
states
\be
\label{12}
 \hat\rho_{sep} = |\; \vp_{sep}\; \rgl \lgl \; \vp_{sep} \; | =
\bigotimes_i \; |\; \vp_i\; \rgl \lgl \; \vp_i \; | \; ,
\ee
while entangled wave functions form entangled states
\be
\label{13}
\hat\rho_{ent} = |\; \vp_{ent}\; \rgl \lgl \; \vp_{ent} \; | = 
\sum_{\al\bt} c_\al^* c_\bt \; 
\bigotimes_i \; |\; \vp_{i\bt}\; \rgl \lgl \; \vp_{i\al} \; | \;  .
\ee
More generally, a state is separable if and only if it has the structure
\be
\label{14}
\hat\rho_{sep} = \sum_\al \lbd_\al \; \bigotimes_i \hat\rho_{i\al} \; ,
\ee
in which 
$$
0 \leq \lbd_\al \leq 1 \; , \qquad \sum_\al \lbd_\al = 1
$$
and $\hat{\rho}_{i \alpha}$ is a statistical operator acting on a partial Hilbert 
space $\mathcal{H}_i$. A state that cannot be represented in the above form is 
entangled.

\subsection{Entangled operators}

The notion of entangled states can be generalized to trace-class entangled 
operators \cite{Yukalov_18,Yukalov_19,Yukalov_20}. Let us consider an algebra 
$\mathcal{A} = \{\hat{A}\}$ of trace-class operators acting on a Hilbert space 
$\mathcal{H}_A$. For any two operators $\hat{A}_1$ and $\hat{A}_2$ from the 
algebra $\mathcal{A}$ one can introduce the Hilbert--Schmidt scalar product
$$
(\hat A_1, \; \hat A_2) \equiv {\rm Tr}_\cH (\hat A_1^+ \hat A_2) \; .
$$
The triple of the operator algebra $\mathcal{A}$, acting on the Hilbert space 
$\mathcal{H}_A$, and the Hilbert--Schmidt scalar product form the Hilbert--Schmidt 
space
$$
\widetilde\cA \equiv \{\cA , \; \cH_A , (, ) \} \;  .
$$
Similarly, one can define the Hilbert--Schmidt space for another trace-class operator 
algebra $\mathcal{B}$ as
$$
\widetilde\cB \equiv \{\cB , \; \cH_B , (, ) \} \; .
$$
 
The composite Hilbert--Schmidt space is given by the tensor product
\be
\label{15}
 \widetilde\cA \bigotimes \widetilde\cB = \{\cA , \; \cH_A , (, ) \}
\bigotimes \{\cB , \; \cH_B , (, ) \} .
\ee
An operator $\hat{C}_{sep}$ from the composite Hilbert--Schmidt space in Equation 
(\ref{15}) is separable if and only if it can be represented in the form
\be
\label{16}
 \hat C_{sep} = \sum_\al \lbd_\al \hat A_\al \bigotimes \hat B_\al \qquad
( \hat A_\al \in \widetilde\cA , \; \hat B_\al \in \widetilde\cB ) \;  .
\ee
Otherwise, the operator is entangled. 

In general, an operator $\hat{C}_{sep}$, defined on the Hilbert space in Equation 
(\ref{6}) is separable if and only if it can be represented as
\be
\label{17}
  \hat C_{sep} = \sum_\al \lbd_\al \bigotimes_i \hat A_{i\al} \;  ,
\ee
where the operators $\hat{A}_{i \alpha}$ act on $\mathcal{H}_i$. 

These definitions lift the notion of entanglement to the operator level. In the 
Hilbert--Schmidt space, operators are isomorphic to the space members, so that the 
distinction between entangled and separable states in the Hilbert--Schmidt space 
becomes similar to that in the Hilbert space 
\cite{Yukalov_18,Yukalov_19,Yukalov_20,Zanardi_26,Balakrishnan_27,Macchiavello_28,Kong_29}.

\section{Entangling operations}

The other notion is the entanglement production by quantum operations. Considering the
operators acting on a Hilbert space $\mathcal{H}$, it is possible to distinguish two 
types of the operator actions, entangling and nonentangling. If an operator $\hat{A}$,
acting on any function from the disentangled set $\mathcal{D}$, leaves the function in 
this set, it is called a nonentangling operator \cite{Gohberg_30,Crouzeux_31},
\be
\label{18}
 \hat A \cD \rightarrow \cD \qquad ( {\rm nonentangling} ) \; .
\ee
However, if there exists at least one function of the disentangled set $\mathcal{D}$
that becomes entangled under the action of the operator, this operator is termed 
entangling \cite{Fan_32,Dao_33}. The strongest type of an entangling operator is a 
universal entangling operator that makes all disentangled functions entangled 
\cite{Chen_34},
\be
\label{19}
 \hat A \cD \rightarrow \cH\setminus \cD \qquad ( {\rm entangling} ) \;  .
\ee

A principal problem is how to measure the entangling power of operators. When one 
is interested in just a few wave functions, it is admissible to analyze the action 
of a given operator on all the functions of interest. In the case of a bipartite 
system, one can check the amount of the produced entanglement by studying the 
entanglement entropy for the considered few wave functions 
\cite{Zanardi_26,Macchiavello_28,Kong_29,Zanardi_35}. This, however, does not allow 
for quantifying the entangling power of the examined operator on the whole Hilbert 
space.  

A general measure of entanglement production for arbitrary trace-class operators 
acting on a Hilbert space was advanced in Refs. \cite{Yukalov_36,Yukalov_37}. The 
idea behind the definition of this measure is to compare the action of the given 
operator $\hat{A}$ with the action of its nonentangling counterpart
\be
\label{20}
\hat A^\otimes \equiv 
\frac{\bigotimes_{i=1}^n \hat A_i}{({\rm Tr}_\cH\hat A)^{n-1} } \;   ,
\ee
in which $\hat{A}_i$ are partially traced operators 
\be
\label{21}
\hat A_i  \equiv {\rm Tr}_{\cH / \cH_i} \hat A \;  .
\ee
The coefficient in Equation (\ref{20}) is defined so that to preserve the trace 
normalization
\be
\label{22}
 {\rm Tr}_\cH \hat A = {\rm Tr}_\cH \hat A^\otimes \; .
\ee

The measure of entanglement production \cite{Yukalov_36,Yukalov_37} by an operator 
$\hat{A}$ on a Hilbert space $\mathcal{H}$ is
\be
\label{23}
 \ep(\hat A) \equiv \log \; 
\frac{||\; \hat A\;||}{||\; \hat A^\otimes \;|| } \;  .
\ee
This measure is based on the comparison of the action of an operator on the whole 
Hilbert space with the action of its nonentangling counterpart that leaves invariant 
the disentangled set,
$$
\hat A \cH \rightarrow \cH \; , \qquad \hat A^\otimes \cD \rightarrow \cD \; .
$$

It is useful to emphasize that, as has been proved 
\cite{Marcus_38,Westwick_39,Beasley_40,Alfsen_41,Johnston_42,Friedland_43}, the 
only operators preserving separability are the operators having the form of tensor 
products of local operators and a swap operator permuting Hilbert subspaces in the 
tensor product of the total Hilbert space of a composite system. However, the action 
of the swap operator is trivial, merely permuting the indices labeling the subspaces. 
Up to the enumeration of subspaces, the product operators are the sole operators 
preserving the separability of functions. The tensor-product operators, as is evident, 
do not produce entanglement,
$$
 \ep(\hat A^\otimes ) = 0 \; .
$$

Thus, the entanglement-production measure (\ref{23}) is zero for nonentangling 
operators, and also it is continuous, additive, and invariant under local unitary 
operations \cite{Yukalov_37,Yukalov_44}. As it should be for being a measure, it 
is semipositive. The sketch of the proof of this important property is as follows 
\cite{Yukalov_37,Yukalov_44}. 

The set of trace-class operators $\hat{A}$ acting on the Hilbert space $\cH$, with 
a given operator norm, forms the Banach space
\be
\label{24}
\cB(\hat A) = \{\hat A , \; \cH , \; ||\; \hat A\; || \} \; ,
\ee 
which is a complete normed linear space. Similarly, the set of the product operators 
$\hat{A}^\otimes$, leaving invariant the disentangled set $\mathcal{D}$, composes 
the Banach space
\be
\label{25}
 \cB(\hat A^\otimes) = 
\{\hat A^\otimes , \; \cD , \; ||\; \hat A^\otimes\; || \} \;  .
\ee
The latter space, by definition, is a subspace of the Banach space seen in 
Equation (\ref{24}), 
$$
 \cB(\hat A^\otimes) \subset \cB(\hat A) \;  .
$$
Then it is admissible to define a projector transforming the members of space in 
Equation (\ref{24}) into the members of space in Equation (\ref{25}),
\be
\label{26}
 \hat P_\otimes \hat A = \hat A^\otimes \;  ,
\ee
with the standard projector properties
$$
 \hat P_\otimes^2 =  \hat P_\otimes \; , \qquad  
\hat P_\otimes^+ = \hat P_\otimes \; , \qquad 
|| \; \hat P_\otimes \; || = 1 \; . 
$$
Therefore we have
$$
|| \; \hat A^\otimes \; || = || \; \hat P_\otimes \hat A \; || \leq
|| \; \hat P_\otimes \; || \cdot ||\; \hat A \; || = ||\; \hat A \; || \; ,
$$
from where the semi-positivity of the measure follows:
\be
\label{27}
 \ep(\hat A ) \geq 0 \;  .
\ee
 
It is important to stress that the entangled structure of an operator and entanglement 
production by this operator are quite different notions. An operator can be separable 
but entangling. Thus, the action of the separable operator in Equation (\ref{17}) on 
the separable function in Equation (\ref{7}) results in an entangled function,
$$
 \hat C_{sep} \vp_{sep} = \vp_{ent} \;  ,
$$
where
$$
 \vp_{ent} = \sum_\al \lbd_\al \bigotimes_i \vp_{i\al}' \qquad
(\vp_{i\al}'\equiv \hat A_{i\al}\vp_i ) \; .
$$

It is possible to notice that the measure of entanglement production in Equation 
(\ref{23}) and the order indices in Equation (\ref{2}), although having rather 
different meanings, but are connected with each other through the relations
$$
 \ep(\hat A ) = \log\; 
\frac{|\; {\rm Tr}_\cH\hat A\;|^{\om(\hat A)} }{||\; \hat A^\otimes \; || } \; ,
\qquad
\om(\hat A) = 
\frac{\ep(\hat A) + \log||\hat A^\otimes||}{\log|{\rm Tr}_\cH\hat A|} \; .
$$
The relations between these quantities will be considered in more detail below.

\section{Density matrices}

To illustrate physical applications of the introduced notions, it is reasonable to 
consider such important physical quantities as reduced density matrices, which can 
be treated as matrix elements of reduced density operators \cite{Coleman_45}. For 
instance, the first-order density operator
\be
\label{28}
\hat\rho_1 = [\; \rho(x,x') \; ]
\ee
is expressed through the matrix elements
\be
\label{29}
 \rho(x,x') = {\rm Tr}_\cF \psi(x) \; \hat\rho \; \hat\psi^\dgr(x') =
 \lgl \; \psi^\dgr(x') \psi(x) \; \rgl \; .
\ee
Here $x$ denotes a set of variables, such as spatial coordinates and spin, $\psi(x)$ 
are field operators, the trace is over the Fock space generated by the field operators 
\cite{Yukalov_46}, and $\hat{\rho}$ is a statistical operator. The second-order 
density operator
\be
\label{30}
\hat\rho_2 = [\; \rho_2(x_1,x_2,x_1',x_2') \; ]
\ee
has the matrix elements
$$
\rho_2(x_1,x_2,x_1',x_2') = {\rm Tr}_\cF \psi(x_1) \psi(x_2) \; \hat\rho \;
\psi^\dgr(x_2') \psi^\dgr(x_1') =
$$
\be
\label{31}
=  
\lgl \; \psi^\dgr(x_2') \psi^\dgr(x_1') \psi(x_1) \psi(x_2)\; \rgl \; .
\ee
Generally, the $n$-th order density operator is defined through the matrix elements
$$
\rho_n(x_1,x_2,\ldots,x_n,x_1',x_2',\ldots,x_n') = {\rm Tr}_\cF 
\psi(x_1) \psi(x_2) \ldots \psi(x_n) \; \hat\rho \;
\psi^\dgr(x_n') \ldots \psi^\dgr(x_2') \psi^\dgr(x_1') =
$$
\be
\label{32}
=
\lgl \; \psi^\dgr(x_n') \ldots \psi^\dgr(x_2') \psi^\dgr(x_1')
\psi(x_1) \psi(x_2) \ldots \psi(x_n) \; \rgl \; .
\ee

The reduced density matrices of different orders are connected with each other by 
means of the relations
$$
\int \rho_n(x_1,x_2,\ldots,x_n,x_1',x_2',\ldots,x_n)\; dx_n =
$$
\be
\label{33}
 =
(N - n + 1) \rho_{n-1}(x_1,x_2,\ldots,x_{n-1},x_1',x_2',\ldots,x_{n-1}')  \;  ,
\ee
in which $N$ is the number of particles in the system. For example,
\be
\label{34}
 \int \rho_2(x_1,x_2,x_1',x_2)\; dx_2 =  ( N - 1 ) \rho(x_1,x_1') \; .
\ee
The trace operation here implies the summation over the variable $x$, defined as
\be
\label{35}
 {\rm Tr}\hat\rho_n =  \int \rho_n(x_1,x_2,\ldots,x_n,x_1,x_2,\ldots,x_n)\; 
dx_1 dx_2 \ldots dx_n  = \frac{N!}{(N-n)!} \; .
\ee
In particular
\be
\label{36}
 {\rm Tr}\hat\rho_1 =  \int \rho(x,x) \; dx = N \; , \qquad
 {\rm Tr}\hat\rho_2 =  \int \rho_2(x_1,x_2,x_1,x_2) \; dx_1 dx_2 = N ( N - 1 ) \; .
\ee
The non-entangling product operator for the second-order density matrix is expressed
through the matrix elements
\be
\label{37}
 \rho_2^\otimes(x_1,x_2,x_1',x_2') = 
\frac{N-1}{N} \; \rho(x_1,x_1') \rho(x_2,x_2') \; ,
\ee
so that
\be
\label{38}
 ||\;\hat\rho_2^\otimes\;|| = \frac{N-1}{N} \; ||\;\hat\rho_1\;||^2 \; .
\ee

Keeping in mind the operator norm in Equation (\ref{5}), associated with the vector 
norm, we have
$$
||\;\hat\rho_1\;|| = \sup_k (\vp_k , \; \hat\rho_1\vp_k) \qquad 
( ||\;\vp_k\;|| = 1 ) \; ,
$$
\be
\label{39}
 ||\;\hat\rho_2\;|| = \sup_{kp} (\vp_k \vp_p, \; \hat\rho_2\vp_k \vp_p) \; ,
\ee
where $\{\varphi_k\}$ is a natural ortho-normalized basis \cite{Coleman_45} and $k$ 
is a multi-index labeling quantum states. In first order, we have
\be
\label{40}
 ||\;\hat\rho_1\;|| = \sup_k N_k \;  ,
\ee
with the notation
$$
 N_k = \int \vp_k^*(x) \rho(x,x') \vp_k(x') \; dx dx' \;  .
$$
In the second order, we find
\be
\label{41}
 ||\;\hat\rho_2\;|| = \sup_{kp} N_{kp} \;  ,
\ee
with
$$
N_{kp} = \int \vp_k^*(x_1) \vp_p^*(x_2) \rho_2(x_1,x_2,x_1',x_2') 
\vp_p(x_2') \vp_k(x_1') \;dx_1 dx_2 dx_1' dx_2' \;  .
$$

In this way, we can calculate the order indices of density matrices
\be
\label{42}
\om(\rho_n) = \frac{\log||\;\hat\rho_n\;||}{\log|\;{\rm Tr}\hat\rho_n\;|}
\ee
as well as the measure of entanglement production
\be
\label{43}
 \ep(\hat\rho_n) = 
\log\; \frac{||\;\hat\rho_n\;||}{||\;\hat\rho_n^\otimes\;||} \; .
\ee

Using the properties of reduced density matrices, one can show 
\cite{Coleman_9,Yukalov_10,Coleman_45,Yang_47} that the order indices, depending on 
statistics, satisfy the inequalities
\be
\label{44}
\om(\hat\rho_n) \leq 1 \qquad ( {\rm Bose} )
\ee
for Bose--Einstein statistics and
\be
\label{45}
\om(\hat\rho_{2n}) \leq \frac{1}{2} \; , \qquad
\om(\hat\rho_{2n+1}) \leq \frac{n}{2n+1} \qquad ( {\rm Fermi} )
\ee
for Fermi--Dirac statistics. When the upper boundary is reached, this signifies the 
occurrence of long-range order. Otherwise, there can only be mid-range or short-range 
order.

\section{Correlation matrices}

Reduced density matrices are a particular case of correlation functions constructed 
by means of the field operators. In general, it is possible to consider correlation 
functions based on some other operators. Thus, taking an arbitrary operator 
$\hat A(x)$, acting on the Fock space, one can set the correlation functions 
$$
C_n(x_1,x_2,\ldots,x_n,x_1',x_2',\ldots,x_n')  \equiv 
{\rm Tr}_\cF \hat A(x_1) \hat A(x_2) \ldots \hat A(x_n) \; \hat\rho \;
\hat A(x_n') \ldots \hat A(x_2') \hat A(x_1') =
$$
\be
\label{46}
= 
\lgl \; \hat A(x_n') \ldots \hat A(x_2') \hat A(x_1')  
\hat A(x_1) \hat A(x_2) \ldots \hat A(x_n) \; \rgl \; .
\ee
Then, one can introduce the correlation operator
\be
\label{47}
 \hat C_n = [\; C_n(x_1,x_2,\ldots,x_n,x_1',x_2',\ldots,x_n') \; ] \;  ,
\ee
whose matrix elements are the above correlation functions. For the correlation 
operator, it is straightforward to define the order indices
\be
\label{48}
\om(\hat C_n) = \frac{\log||\;\hat C_n\;||}{\log|\;{\rm Tr}\hat C_n\;|} 
\ee
and the entanglement-production measure
\be
\label{49}
 \ep(\hat C_n) = \log\; \frac{||\;\hat C_n\;||}{||\;\hat C_n^\otimes\;||} \; .
\ee
By choosing appropriate correlation functions one can quantify the properties of 
arbitrary physical systems.

\section{Examples of order indices}

Order indices are defined by the structure of the system state and can essentially 
vary under phase transformations \cite{Coleman_63}. Some examples of order indices 
that have been considered in literature are mentioned below. 

\subsection{Superconducting state}

The structure of reduced density matrices have been analyzed in several works 
\cite{Coleman_6,Coleman_8,Coleman_9,Coleman_45,Yang_47}. In the thermodynamic limit, 
the order indices of density matrices in a three-dimensional system are zero for the 
normal state and take the values
\be
\label{50}
 \om(\hat\rho_{2n}) = \frac{1}{2} \; , \qquad  
\om(\hat\rho_{2n+1}) = \frac{n}{2n+1}  \; ,
\ee
for superconducting state. This corresponds to even long-range order.

\subsection{Bose-condensed system}

Under the usual Bose--Einstein condensation into the state with zero momentum, in the 
thermodynamic limit in three dimensions, all order indices grow from zero to one,
becoming
\be
\label{51}
 \om(\hat\rho_{n}) = 1 \; .
\ee
This corresponds to the total long-range order \cite{Coleman_45,Yang_47}. Developing
long-range order strongly influences the structure of correlation functions and density 
matrices \cite{Yukalov_64,Cummings_65,Ghassib_66,Chow_67}.

\subsection{Even Bose condensate}

There are theoretical speculations \cite{Valatin_48,Girardeau_49,Coniglio_50,Evans_51,
Coniglio_52,Hasting_53,Kondratenko_54,Peletminskii_55,Pashitskii_56} that in Bose 
systems there can develop the so-called even condensate, with the formation of pairs 
similar to those for Fermi systems in the superconducting state, when
\be
\label{52}
 \om(\hat\rho_{2n}) = \frac{1}{2} \;  ,
\ee
thus exhibiting even long-range order.

\subsection{Finite-momentum condensate}

If the condensed state is characterized not by zero momentum but by a momentum with 
finite absolute value and random direction 
\cite{Yukalov_64,Yukalov_57,Yukalov_58,Yukalov_59}, then the system possesses 
mid-range order with the order indices that, depending on dimensionality $d$, are
\be
\label{53}
\om(\hat\rho_n) = \frac{1}{d} \;   .
\ee
For $d>1$ this implies a mid-range order \cite{Yukalov_10}.

\subsection{Two-dimensional systems}

Two-dimensional systems below the Berezinskii--Kosterlitz--Thouless transition 
\cite{Berezinskii_4,Kosterlitz_5} temperature $T_K$ possess the order indices 
\cite{Coleman_6}
\be
\label{54}
\om(\hat\rho_n) = 1 \; - \; \frac{\eta}{4} 
\qquad 
( d = 2 , \; T < T_K ) \; ,
\ee
where $\eta$ is the exponent describing the behavior of the pair correlation function 
at large distance. In that case 
$$
\frac{1}{4}  \leq \eta \leq \frac{1}{3} \; ,
$$
so that there exists a mid-range order with the order indices 
$$
\frac{11}{12}  \leq \om(\hat\rho_n) \leq \frac{15}{16} 
\qquad 
( d = 2 , \; T < T_K ) \; .
$$    

Mid-range order also develops in two-dimensional antiferromagnets in a strong magnetic 
field \cite{Gluzman_68,Gluzman_69}.

\subsection{Critical point}

At the point of a second-order phase transition, we have \cite{Coleman_6}
\be
\label{55}
 \om(\hat\rho_n) =  \frac{d+2-\eta}{2d} \qquad ( T = T_c) \; ,
\ee
with $\eta$ being the exponent characterizing the pair correlation function. For 
different dimensions, this gives
$$
 \om(\hat\rho_n) =  \frac{4-\eta}{4} \qquad ( d = 2 , \; T = T_c) \; ,
$$
\be
\label{56}
\om(\hat\rho_n) =  \frac{5-\eta}{6} \qquad ( d = 3 , \; T = T_c) \;   ,
\ee
which implies mid-range order. In particular, for the two-dimensional Ising model, 
we have \cite{Pelisetto_60,Yukalov_61,Yukalov_62} $\eta = 1/4$, hence 
\be
\label{57}
 \om(\hat\rho_n) =  \frac{15}{16} \qquad ( d = 2 ,\; {\rm Ising} , \; T = T_c) \;  .
\ee
For the three-dimensional Heisenberg model, $\eta = 0.036$, therefore
\be
\label{58}
\om(\hat\rho_n) = 0.827 \qquad ( d = 3 ,\; {\rm Heisenberg} , \; T = T_c) \; .
\ee

\subsection{Tonks--Girardeau gas}

One-dimensional systems usually do not exhibit long-range order, although they can 
possess mid-range order. As an example, let us consider the Tonks--Girardeau gas 
\cite{Girardeau_70}. This is a one-dimensional system of impenetrable bosons described 
by the Hamiltonian
\be
\label{59}
\hat H = -\;\frac{1}{2m} \sum_{i=1}^N \frac{\prt^2}{\prt x_i^2}  \; +  \;
\sum_{i=1}^N U(x_i) \;   ,
\ee  
where $U(x)$ is an external potential, and complimented by the condition on the wave 
function
\be
\label{60}
 \psi(x_1,x_2,\ldots,x_N,t) = 0 \qquad ( | \; x_i - x_j \; | \leq a_0 ) \; ,
\ee
where $a_0$ is an effective diameter of particles. Studying the properties of reduced 
density matrices \cite{Lenard_71,Forrester_72,Yukalov_73,Colcelli_74} one concludes 
that the order indices are
\be
\label{61}
 \om(\hat\rho_n) = \frac{1}{2}   
\ee
for both homogeneous as well as for systems trapped in power-law potentials. This 
shows that the Tonks--Girardeau gas possesses a mid-range order.

\subsection{Lieb--Liniger model}

This is a one-dimensional model described by the Hamiltonian \cite{Lieb_75}
\be
\label{62}
\hat H = -\; \frac{1}{2m} \sum_{i=1}^N \frac{\prt^2}{\prt x_i^2} + 
\Phi_0 \sum_{i\neq j} \dlt( x_i - x_j )   ,   
\ee
with point-like interactions of finite strength $\Phi_0$. The characteristic 
dimensionless interaction parameter is defined by the ratio of the typical potential 
energy to the typical kinetic energy
$$
E_K = \frac{\rho^2}{2m} \qquad \left( \rho \equiv \frac{N}{L} \right) \;  ,
$$
which gives
\be
\label{63}
 \gm \equiv \frac{\rho\Phi_0}{E_K} = \frac{2m}{\rho}\;\Phi_0 \;  .
\ee
The model is exactly solvable by Bethe ansatz \cite{Lieb_75,Yang_76}, which 
provides the exact expression of the many-body eigenfunctions \cite{Gaudin_77}. 
At zero temperature, the ground-state energy, the sound velocity $c$, and other 
equilibrium quantities can be expressed in terms of the solution of the Lieb--Liniger 
integral equations \cite{Lieb_75}. The other useful quantity is the Luttinger 
parameter \cite{Giamarchi_78} 
\be
\label{64}
 M \equiv \frac{v_F}{c} \qquad \left( v_F \equiv \frac{\pi\rho}{m} \right) \;  ,
\ee
in which $v_F$ is the Fermi velocity and $c$ is the sound velocity. Note that this 
parameter reminds the effective Mach number used for characterizing statistical 
systems \cite{Yukalov_79}. Solving the Lieb--Liniger integral equations one has 
access to the sound velocity $c$ and the Luttinger parameter $M$ for any values 
of the coupling constant \cite{Cazalilla_80,Lang_81,Lang_82,Ristivojevic_83}. 
 
The order index of the first-order density matrix can be estimated \cite{Colcelli_84} 
as
\be
\label{65}
\om(\hat\rho_1) = 1 \; - \; \frac{1}{2M}\;   .
\ee
For small $\gamma$, one has
$$
c \simeq \frac{\rho}{m} \; \sqrt{\gm} \; , \qquad 
M \simeq \frac{\pi}{\sqrt{\gm}} \qquad ( \gm \ra 0 ) \;  ,
$$
so that 
\be
\label{66}
\om(\hat\rho_1) \simeq 1\; - \; \frac{\sqrt{\gm}}{2\pi} \qquad ( \gm \ra 0 ) \; .
\ee
While for strong interactions, when
$$
c \simeq v_F \; , \qquad  M \simeq 1 \qquad ( \gm \ra \infty ) \; ,
$$
the order index tends to that of the Tonks--Girardeau gas,
\be
\label{67}
\om(\hat\rho_1) \simeq  \frac{1}{2} \qquad ( \gm \ra \infty ) \; .
\ee
As we see, the Lieb--Liniger model demonstrates a mid-range order. Formally, the 
order index tends to one in the limit of the ideal gas, when $\gamma=0$. However, 
the ideal one-dimensional uniform gas is unstable \cite{Yukalov_85,Yukalov_86}.

\subsection{Finite systems}

Strictly speaking, usually long-range order can arise only in thermodynamic limit. 
In finite systems, order parameters not always can be rigorously defined. However, 
the order indices can be well defined for any finite system. It is reasonable to 
expect that in finite systems, instead of long-range order, there could exist only 
mid-range or short-range order. In that case, the order indices vary in the interval
$$
0 \leq \om(\hat\rho_n) < 1 \qquad ({\rm Bose} ) \; ,
$$
\be
\label{68}
0 \leq \om(\hat\rho_n) < \frac{1}{2} \qquad ({\rm Fermi} ) \; ,
\ee
depending on the number of particles in the system, on the interaction strength, and 
on other system parameters \cite{Yukalov_87,Yukalov_88}. We shall discuss the details 
of calculating the order indices for systems with a finite, although large, number of 
particles in Secs. 9 and 10.

\section{Examples of entanglement production}

Now we shall give several examples of the entanglement-production measure calculated 
\cite{Yukalov_37,Yukalov_89} according to Sec. 4. For bipartite states, it is also 
possible to find the entanglement entropy
\be
\label{69}
 S(\hat\rho_i) = - {\rm Tr}_{\cH_i} \hat\rho_i \ln \hat\rho_i \;  ,
\ee
where
$$
\hat\rho_i =  {\rm Tr}_{\cH/\cH_i} \hat\rho   
$$
is a partially traced statistical operator. The latter can be compared with the 
entanglement-production measure in Equation (\ref{23}). In general, measures in 
Equations (\ref{23}) and (\ref{69}) do not need to be equal. In addition, the 
entanglement entropy in Equation (\ref{69}) can be defined only for bipartite 
states, while the entanglement-production measure in Equation (\ref{23}) can be 
defined for arbitrary states and operators.

\subsection{Einstein--Podolsky--Rosen-states}

The statistical operator of this entangled pure state is
$$
\hat\rho_{EPR} = |\; EPR \; \rgl \lgl \; EPR \; |  \; ,
$$
in which
$$
|\; EPR \; \rgl = \frac{1}{\sqrt{2} } \; 
(\; |\; 12 \; \rgl \; \pm \; |\; 21 \; \rgl \; ) \; .
$$   
This is a two-particle two-mode state. For the entanglement-production measure, we 
have
\be
\label{70}
 \ep(\hat\rho_{EPR}) = \log 2 \;  .
\ee
This coincides with the entanglement entropy in Equation (\ref{69}).

\subsection{Bell states}

This is also an entangled two-particle two-mode pure state with the statistical 
operator
$$
\hat\rho_B = | B \rgl \lgl B | \;  ,
$$
where
$$
| B \rgl = \frac{1}{\sqrt{2}} \; 
(\; | \; 11 \; \rgl \; \pm \; | \; 22 \; \rgl \; ) \;  .
$$
The measure in Equation (\ref{23}) 
\be
\label{71}
\ep(\hat\rho_B) = \log 2
\ee
equals the entanglement entropy in Equation (\ref{69}).

\subsection{Greenberger--Horne--Zeilinger states}

This is an $N$-particle two-mode state
$$
\hat\rho_{GHZ} = |\; GHZ \; \rgl \lgl \; GHZ \; |  \;   ,
$$
with
$$
  |\; GHZ \; \rgl = \frac{1}{\sqrt{2}} \; 
( \; |\; 11 \ldots 1 \; \rgl \; \pm \; | \; 22\ldots 2 \; \rgl \; ) \;  .
$$
For the measure in Equation (\ref{23}), we get
\be
\label{72}
\ep(\hat\rho_{GHZ}) = ( N - 1 ) \log 2 \; .
\ee
For the number of particles more than two, the entanglement entropy is not defined.

\subsection{Multicat states}

The statistical operator for the two-mode $N$-particle state
$$
\hat\rho_{MC} = |\; MC \; \rgl \lgl \; MC \; |
$$
is expressed through the multicat function
$$
|\; MC \; \rgl =  c_1 |\; 11 \ldots 1 \; \rgl \; + \; c_2 |\; 22\ldots 2 \; \rgl \; ,
$$
where
$$
 | c_1 |^2  + | c_2 |^2 = 1 \; .
$$
The entanglement-production measure becomes
\be
\label{73}
 \ep(\hat\rho_{MC} ) = ( 1 - N ) \log \sup \{ | c_1 |^2 , \; | c_2 |^2 | \} \; .
\ee
Multicat states can be realized in multiparticle systems where each particle 
possesses two internal energy states, for instance, for trapped ions subject to 
the action of resonant laser beams \cite{Molmer_90,Sorensen_91} or for Bose-condensed 
neutral atoms under coherent Raman scattering \cite{Helmerson_92}. Instead of internal 
single-particle states, one can use collective nonlinear states created by means of 
the resonant excitation of topological coherent modes in trapped Bose-Einstein 
condensates \cite{Yukalov_93,Yukalov_94,Yukalov_95,Yukalov_96}.

\subsection{Multimode states}

The state of $N$ particles possessing $N_M$ modes each,
$$
\hat\rho_{MM} = |\; MM \; \rgl \lgl \; MM \; |  \;   ,   
$$
is formed by the multimode function
$$
 |\; MM \; \rgl = \sum_{n=1}^{N_M} c_n | nn\ldots n \rgl \;  ,
$$
for which
$$
 \sum_{n=1}^{N_M} | c_n|^2 = 1 \; .
$$
This is a generalization of the multicat state to the case of $N$ particles. The measure 
of entanglement production is
\be
\label{74}
\ep(\hat\rho_{MM}) = ( 1 - N ) \log \sup_n | c_n|^2 \; .
\ee
The multimode functions can be represented by coherent states \cite{Carusotto_97}.

\subsection{Hatree--Fock states}

The state with the statistical operator
$$
\hat\rho_{HF} = |\; HF \; \rgl \lgl \; HF \; |  \; 
$$ 
is formed by $N$-particle symmetrized or antisymmetrized functions
$$
|\; HF \; \rgl = \frac{1}{\sqrt{N!} } \sum_{sym}  | 1 2 \ldots N \rgl \;   .
$$
Such functions represent $N$ identical particles, bosons or fermions, in $N$ 
different energy states. The entanglement-production measure reads as
\be
\label{75}
\ep(\hat\rho_{HF}) =  \log \; \frac{N^N}{N!} \;   .
\ee
In the case of two particles, the measure reduces to the that for the 
Einstein--Podolsky--Rosen state,
\be
\label{76}
\ep(\hat\rho_{HF}) = \log 2 \qquad ( N = 2 ) \;  .
\ee
For macroscopic systems, it tends to the expression
\be
\label{77}
\ep(\hat\rho_{HF}) \simeq N \log e \qquad ( N \gg 1 ) \; .
\ee

\subsection{Reduced statistical operators}

In the above examples, pure states are considered. The measure of entanglement production, 
being general, can be found for mixed states as well. To this end, we can study reduced 
statistical operators defined through partial traces of the Hartree--Fock state
$$
\hat\rho_n \equiv {\rm Tr}_{\cH_{n+1}}  {\rm Tr}_{\cH_{n+2}} \ldots
 {\rm Tr}_{\cH_N} \; \hat\rho_{HF} \; ,
$$
with $n = 1,2,\ldots,N-1$. The entanglement-production measure for this operator is
\be
\label{78}
 \ep(\hat\rho_n) = \log\; \frac{(N-n)!N^n}{N!} \; .
\ee
For a large number of particles, this yields
\be
\label{79}
  \ep(\hat\rho_n) \simeq \frac{n(n-1)}{2N}\; \log e \qquad ( N \gg 1 ) \;  .
\ee

\subsection{Gibbs states}

An equilibrium system of $N$ particles, characterized by a Hamiltonian $H$ acting 
on a Hilbert space
$$
 \cH = \bigotimes_{i=1}^N \cH_i \;  ,
$$
is described by the Gibbs statistical operator
$$
\hat\rho = \frac{1}{Z} \; e^{-\bt H} \qquad 
\left( Z \equiv {\rm Tr}_\cH e^{-\bt H} \right) \;  ,
$$
in which $\beta \equiv 1/T$ is inverse temperature. The entanglement-production 
measure can be calculated as is explained in Sec. 4. For two-particle registers, 
represented by Ising and Heisenberg Hamiltonians, this was done in 
Refs. \cite{Yukalov_37,Yukalov_89}.

\section{Werner operator}

The case of the Werner operator \cite{Werner_98} is interesting since, depending on 
parametrization, it can represent a separable or entangled state. The operator has 
the form
\be
\label{80}
 \hat\rho_W = \frac{1}{(d^2-1)d} \; \left[ \; ( d - \gm ) \hat 1 + 
( \gm d - 1)\hat\sgm \; \right] \; ,
\ee
with the unity operator 
$$
 \hat 1 = \sum_{mn} |\; mn \; \rgl \lgl \; nm \; | = 
\hat 1_A \bigotimes \hat 1_B \; ,
$$
where
$$
\hat 1_A = \sum_m  |\; m \; \rgl \lgl \; m \; | \; , \qquad
\hat 1_B = \sum_n  |\; n \; \rgl \lgl \; n \; | \; ,
$$
and the flip operator
$$
 \hat\sgm = \sum_{mn} |\; mn \; \rgl \lgl \; mn \; | = 
\sum_{mn} |\; m \; \rgl \lgl \; n \; |  \bigotimes 
|\; n \; \rgl \lgl \; m \; | \; .
$$
The name of the latter comes from the property of the flip operator to flip 
functions:
$$
\hat\sgm \; | \; mn \; \rgl =  | \; nm \; \rgl \; .
$$
The parametrization of the Werner operator is done by the parameter $\gamma$ defined 
by the equation
$$
\gm = {\rm Tr}_\cH \hat\rho_W \hat \sgm \qquad ( {\rm Tr}_\cH \hat\rho_W= 1 ) \; .
$$

The partially traced operators are
$$
\hat\rho_1 = {\rm Tr}_{\cH_2} \hat\rho_W = \frac{1}{d} \; \hat 1_A \; , 
\qquad
\hat\rho_2 = {\rm Tr}_{\cH_1} \hat\rho_W = \frac{1}{d} \; \hat 1_B \;  .
$$
These reduced operators are normalized to one,
$$
 {\rm Tr}_{\cH_1} \hat\rho_1 = {\rm Tr}_{\cH_2} \hat\rho_2 = 1 \; .
$$
The nonentangling counterpart of the Werner operator is
$$
\hat\rho_W^\otimes = \hat\rho_1 \bigotimes \hat\rho_2 = \frac{1}{d^2}\; \hat 1 \; ,
$$
which gives its norm
$$ 
|| \hat\rho_W^\otimes|| = \frac{1}{d^2} \;    .
$$
In view of the matrix elements
$$
\lgl \; mn \; | \hat 1 \; | \; nm \; \rgl = 1 \; , \qquad
\lgl \; mn \; | \hat\sgm \; | \; nm \; \rgl = \dlt_{mn} \; ,
$$
we have
$$
\lgl \; mn \; | \hat\rho_W \; | \; nm \; \rgl = 
\frac{d-\gm+(\gm d-1)\dlt_{mn}}{(d^2-1)d} \;  .
$$
Thence the norm of the Werner operator is
$$
||\hat\rho_W || = \frac{1}{(d^2 - 1)d} \sup \{ d - \gm, \; ( 1 + \gm )( d - 1 ) \} \;   .
$$
Therefore for the entanglement-production measure we find
\be
\label{81}
\ep(\hat\rho_W) = 
\log [\; \frac{d}{d^2 - 1} \sup\{ d - \gm,\; ( 1 + \gm ) ( d - 1 ) \} \; ] \; .
\ee
The measure is positive for all values of $\gamma$, which means that the Werner 
operator is always entangling. At the same time, the positive partial transpose 
criterion \cite{Peres_99,Horodecki_100} tells us that the Werner state is separable 
if and only if $\gm\geq 0$, while it is entangled for $\gm<0$. However, even being 
separable, the Werner operator is entangling, that is, producing entanglement.

The Werner operator is an explicit example of an operator that can be separable, 
although being entangling, as is discussed in Sec. 4. The entanglement entropy for 
the Werner operator is
$$
 S_j = - {\rm Tr}_{\cH_j} \hat\rho_j \ln\hat\rho_j = \frac{1}{d} \; \ln d \;   .
$$

\section{Bose--Einstein condensation}

The aim of the present section is twofold. First, it shows explicitly how the order 
indices and the entanglement-production measure can be calculated for a system with 
a phase transformation for a large, however finite, number of particles. Second, it
is demonstrated that both these characteristics, order indices as well as entanglement 
production, simultaneously change under phase transitions.     

The correct description of a Bose-condensed system requires the validity of the 
following important stipulations. (i) The theory has to respect conservation laws 
and thermodynamic relations. (ii) The excitation spectrum must be gapless. (iii) The 
Bose--Einstein condensation has to be a phase transition of second order. (iv) The 
system must be stable, satisfying the necessary stability conditions. (v) Reasonable 
quantitative agreement with experiments is required. The validity of these stipulations 
can be achieved in the self-consistent approach based on representative statistical 
ensembles \cite{Yukalov_101,Yukalov_102,Yukalov_103,Yukalov_104} which we use here. 

In the presence of Bose--Einstein condensate, the boson field operators acquire the 
Bogolubov shift \cite{Bogolubov_105,Bogolubov_106}, being represented as the sum
\be
\label{82}
\hat\psi(\br) = \eta(\br) + \psi_1(\br) \;  ,
\ee
in which $\eta$ is the condensate wave function and $\psi_1$ is the field operator of
uncondensed particles. The field operators are also functions of time, which are not 
shown explicitly for the sake of notation compactness. The functional variables $\eta$ 
and $\psi$ are orthogonal with each other,
\be
\label{83}
 \int \eta^*(\br) \psi_1(\br) \; d\br = 0  
\ee
and satisfy the conditions
\be
\label{84} 
\eta(\br) = \lgl \; \hat\psi(\br) \; \rgl \; , \qquad
 \lgl \; \hat\psi_1(\br) \; \rgl = 0 \; .
\ee

The condensate function is normalized to the number of condensed particles,
\be
\label{85}
 N_0 = \int |\; \eta(\br) \; |^2 \; d\br  \;  ,
\ee
while the number of uncondensed particles is
\be
\label{86}
N_1 = \int \lgl \; \hat\psi_1^\dgr(\br) \psi_1(\br) \; \rgl \; d\br  \; .
\ee

The first-order density matrix reads as
\be
\label{87}
\rho(\br,\br') = \eta^*(\br') \eta(\br) + 
\lgl \; \hat\psi_1^\dgr(\br') \psi_1(\br) \; \rgl  \; .
\ee
Its eigenvalues can be represented through the integral
\be
\label{88}
N_k = \int \vp_k^*(\br) \rho(\br,\br') \vp_k(\br') \; d\br  d\br' \;  ,
\ee
in which $\varphi_k$ are the natural orbitals \cite{Coleman_45}. Taking account of 
Equation (\ref{87}) leads to the sum
\be
\label{89}
 N_k = N_{0k} + n_k \;  ,
\ee
where
$$
 N_{0k} \equiv \left| \; \int \eta^*(\br) \vp_k(\br) \; d\br \; \right|^2  
$$
and
$$
 n_k \equiv \lgl \; a_k^\dgr a_k \; \rgl \; , \qquad
a_k \equiv \int \vp_k^*(\br) \psi_1(\br) \; d\br \; .
$$
Thus we obtain the norm
\be
\label{90}
 ||\; \hat\rho_1 \; || =\sup_k ( N_{0k} + n_k ) \;  .
\ee
In the absence of the condensate, $\eta=0$ and $N_{0k}=0$. 

In the presence of the condensate, there also exists the so-called anomalous average
\be
\label{91}
 \sgm_1(\br_1,\br_2) \equiv 
\lgl \; \psi_1(\br_2) \psi_1(\br_1) \; \rgl =
 \sgm_1(\br_2,\br_1)  \; ,
\ee
with the property
$$
 \sgm_1^*(\br_1,\br_2) \equiv 
\lgl \; \psi_1^\dgr(\br_1) \psi_1^\dgr(\br_2) \; \rgl =
 \sgm_1^*(\br_2,\br_1)  \;  .
$$

If the system is uniform, then
\be
\label{92}
\eta(\br) = \sqrt{\rho_0} \qquad \left( \rho_0 \equiv \frac{N_0}{V} \right)
\ee
and it follows 
\be
\label{93}
  N_{0k} = N_0 \dlt_{k0} \; .
\ee

Considering the second-order density matrix, we shall use the simplified notation 
writing just $j$ instead of $\br_j$ and $j'$ instead of $\br_j'$. For instance, 
\be
\label{94}
 \rho_1(1,2) \equiv \rho_1(\br_1,\br_2) = 
\lgl \; \psi_1^\dgr(\br_2) \psi_1(\br_1) \; \rgl \; .
\ee
Then the second-order density matrix reads as
$$
\rho_2(1,2,1',2') = \eta^*(2') \eta^*(1') \eta(1)\eta(2) +  
\rho_1(1,2') \eta^*(1')\eta(2) + \rho_1(1,1') \eta^*(2')\eta(2) +
$$
$$
+ \rho_1(2,2') \eta^*(1')\eta(1) + \rho_1(2,1') \eta^*(2')\eta(1) +
\sgm_1(1,2) \eta^*(2')\eta^*(1') + \sgm_1^*(1',2') \eta(1)\eta(2)  +
$$
$$
+ \lgl \; \psi_1^\dgr(2') \psi_1^\dgr(1') \psi_1(1) \; \rgl \; \eta(2) + 
\lgl \; \psi_1^\dgr(2') \psi_1^\dgr(1') \psi_1(2) \; \rgl \eta(1) +
$$
\be
\label{95}
+ \lgl \; \psi_1^\dgr(2') \psi_1(1) \psi_1(2) \; \rgl \; \eta^*(1') + 
\lgl \; \psi_1^\dgr(1') \psi_1(1) \psi_1(2) \; \rgl \; \eta^*(2') +
\lgl \; \psi_1^\dgr(2') \psi_1^\dgr(1') \psi_1(1) \psi_1(2) \; \rgl \; .
\ee

In what follows, let us consider a uniform system, when Equation (\ref{92}) 
holds true. In that case, the natural orbitals are plane waves. By employing the 
Hartree--Fock--Bogolubov decoupling reduces the second-order density matrix to the 
expression
$$
\rho_2(1,2,1',2') = \rho_0^2 + \rho_0 [\; \rho_1(1,2') + \rho_1(1,1') +
\rho_1(2,2') + \rho_1(2,1') + \sgm_1(1,2) + \sgm_1^*(1',2') \; ] + 
$$
\be
\label{96}
+ \rho_1(1,2') \rho_1(2,1') +
\rho_1(2,2')\rho_1(1,1') + \sgm_1^*(1',2')\sgm_1(1,2) \;  .
\ee

The matrix element in Equation (\ref{94}) becomes
\be
\label{97}
\rho_1(\br,\br') = \frac{1}{V} \sum_{k\neq 0} n_k e^{i\bk\cdot(\br-\br')} \; ,
\ee
where
$$
n_k \equiv \lgl \; a_k^\dgr a_k \; \rgl = 
\frac{1}{V} \int \rho_1(\br',\br) e^{i\bk\cdot(\br-\br')}\; d\br d\br' \; .
$$
Similarly, the anomalous average in Equation (\ref{91}) takes the form
\be
\label{98}
\sgm_1(\br,\br') = \frac{1}{V} \sum_{k\neq 0} \sgm_k e^{i\bk\cdot(\br-\br')} \; ,
\ee
with
$$
\sgm_k \equiv \lgl \; a_k a_{-k} \; \rgl = 
\frac{1}{V} \int \sgm_1(\br',\br) e^{i\bk\cdot(\br-\br')}\; d\br d\br' \;   .
$$
For the norm of the second-order density operator, we get
\be
\label{99}
||\; \hat\rho_2 \; || = \sup_{kp} \left( N_0^2 \dlt_{k0}\dlt_{p0} +
2n_k n_p + \sgm_k \sgm_p \right) \; .
\ee
  
In that way, we find the order indices for the density operators of first order,
\be
\label{100}
 \om(\hat\rho_1) = \frac{\log||\;\hat\rho_1\;||}{\log|\;{\rm Tr}\hat\rho_1\;|} =
\frac{\log\sup_k N_k}{\log N} \;  ,
\ee
in which
$$
 N_k = N_0 \dlt_{k0} + n_k \;  ,
$$
and of second order 
\be 
\label{101}
  \om(\hat\rho_2) = \frac{\log||\;\hat\rho_2\;||}{\log|\;{\rm Tr}\hat\rho_2\;|} =
\frac{\log\sup_{kp} N_{kp}}{2\log N} \;  ,
\ee
where
$$
 N_{kp} = N_0^2 \dlt_{k0}\dlt_{p0} + 2 n_k n_p + \sgm_k \sgm_p \; .
$$
For the entanglement-production measure, we obtain
\be
\label{102}
 \ep(\hat\rho_2) = 
\log\; \frac{||\;\hat\rho_2\;||}{||\;\hat\rho_2^\otimes\;||} = 
\log\; \frac{\sup_{kp}N_{kp}}{(\sup_k N_k)^2} \; .
\ee
These quantities can be calculated \cite{Yukalov_87,Yukalov_88} as functions of the 
number of particles and other system parameters.

For illustration, let us consider the limiting case of $N \ra \infty$. Then, for 
temperatures below the condensation temperature $T_c$, we find that
$$
\sup_k N_k \simeq N_0 \propto N \; , \qquad 
\sup_{kp} N_{kp} \simeq N_0^2 \propto N^2 \qquad ( T < T_c ) \; ,
$$
while $\sup_k n_k \propto N^{1/3}$. At temperatures above the condensation 
temperature $T_c$, we have 
$$
\sup_k N_k = \sup_k n_k \; , \qquad 
\sup_{kp} N_{kp} = 2\sup_k n_k^2 \qquad ( T > T_c ) \; .
$$
Therefore the studied order indices are
\begin{eqnarray}
\label{103}
\om(\hat\rho_1) = \om(\hat\rho_2) = \left\{ \begin{array}{ll}
1 , ~ & ~ T < T_c \\
0 , ~ & ~ T > T_c \end{array}
\right.
\end{eqnarray}
and the entanglement-production measure is
\begin{eqnarray}
\label{104}
\ep(\hat\rho_2) =  \left\{ \begin{array}{ll}
0 , ~ & ~ T < T_c \\
\log 2 , ~ & ~ T > T_c \end{array} \; .
\right.
\end{eqnarray}
At the phase transition point, both these quantities experience noticeable change. 
The appearance of order is accompanied by the reduction of the entanglement production.

\section{Magnetic transitions}

Magnetic systems are characterized by spin operators ${\bf S}_j$ located at lattice 
sites enumerated by $j=1,2,\ldots,N$. Instead of density matrices defined through field 
operators, we need now to introduce correlation matrices, as in Sec. 6, composed of spin 
operators \cite{Yukalov_10}. Thus the first-order spin correlation matrix
\be
\label{105}
\hat C_1 = [\; C_{ij}\; ]
\ee
is composed of the matrix elements 
\be
\label{106}
 C_{ij} =\lgl \; \bS_i \cdot \bS_j \; \rgl \;  .
\ee
The second-order spin correlation matrix
\be
\label{107}
 \hat C_2 = [\; C_{ijmn}\; ]  
\ee
is formed by the correlation functions
\be
\label{108}
 C_{ijmn} =\lgl \; \bS_i (\bS_j \cdot \bS_m) \bS_n \; \rgl \; .
\ee

Keeping in mind that
$$
 \bS_j^2 = S ( S + 1 ) \; ,
$$
we have the traces
\be
\label{109}
{\rm Tr} \hat C_1 = \sum_j C_{jj} = S ( S + 1 ) N
\ee
and 
\be
\label{110}
{\rm Tr} \hat C_2 = \sum_{ij} C_{ijji} = [\; S ( S + 1 ) N \;]^2 \;  .
\ee
   
In calculating the matrix norms, we use the natural lattice orbital
\be
\label{111}
 \vp_k(\ba_j) = \frac{1}{\sqrt{N}} \; e^{i\bk \cdot \ba_j} \; .
\ee
Then the norm of the first-order correlation matrix is
\be
\label{112}
 ||\; \hat C_1 \; || = 
\sup_k \left| \; \sum_{ij} \vp_k^*(\ba_i) C_{ij} \vp_k(\ba_j) \; \right| \;  .
\ee
Employing the properties
$$
 C_{ijjn} =  S ( S + 1 ) C_{in} \; , \qquad 
|\; \vp_k(\ba_j) \; |^2 = \frac{1}{N} \; , 
$$
and separating the terms with coinciding and non-coinciding lattice sites, we get
\be
\label{113}
 ||\; \hat C_1 \; || =  \sup_k \left| \; 
S ( S + 1 ) + \sum_{i\neq j} \vp_k^*(\ba_i) C_{ij} \vp_k(\ba_j) \; \right| \; .
\ee

Similarly, for the second-order spin correlation matrix, we have the norm
\be 
\label{114}
||\; \hat C_2 \; || =  \sup_{kp} \left| \;  \sum_{ijmn} 
\vp_k^*(\ba_i) \vp_p^*(\ba_j) C_{ijmn} \vp_p(\ba_m) \vp_k(\ba_n) \; \right|\; ,
\ee
which leads to 
$$
||\; \hat C_2 \; || =  \sup_{kp} \left| \; 
[\; S ( S + 1 ) \;]^2 + 2S ( S + 1) \sum_{i\neq j} \vp_k^*(\ba_i) C_{ij} \vp_k(\ba_j) 
+ \right.
$$
\be
\label{115}
+ \left.
\sum_{i\neq n} \; \sum_{j\neq m} 
\vp_k^*(\ba_i) \vp_p^*(\ba_j) C_{ijmn} \vp_p(\ba_m) \vp_k(\ba_n) \; \right| \; .
\ee

Defining the magnetization
\be
\label{116}   
 \bM = \lgl \; \bS_j \; \rgl \; ,
\ee
for different lattice sites, one can resort to the mean-field approximation
\be
\label{117}
 \lgl \; \bS_i \cdot \bS_j \; \rgl = 
\lgl \; \bS_i \; \rgl \lgl \; \bS_j \; \rgl = M^2 \qquad ( i \neq j) \; .
\ee
Then we have
\begin{eqnarray}
\nonumber
C_{ij} = \left\{ \begin{array}{ll}
M^2 , ~ & ~ i \neq j \\
S ( S+1 ) ,  ~ & ~ i = j \; ,
\end{array} \right.
\end{eqnarray}
which gives
\be
\label{118}
 || \; \hat C_1 \; || = S ( S + 1 ) + ( N - 1 ) M^2 \; .
\ee
Depending on the number of coinciding and non-coinciding lattice sites, we find
$$
\lgl \; \bS_i (\bS_j \cdot \bS_m) \bS_n \; \rgl =  
\lgl \; \bS_i \cdot \bS_n \; \rgl \lgl \; \bS_j \cdot \bS_m \; \rgl \qquad
( i \neq n , ~ j \neq m , ~  j \neq n , ~ m \neq n ) \; ,
$$
$$
\lgl \; \bS_j (\bS_j \cdot \bS_m) \bS_n \; \rgl =  
\lgl \; \bS_j \cdot \bS_j \; \rgl \lgl \; \bS_m \cdot \bS_n \; \rgl \qquad
( j \neq n , ~ j \neq m , ~  m \neq n ) \; ,
$$
$$
\lgl \; \bS_i (\bS_j \cdot \bS_n) \bS_n \; \rgl =  
\lgl \; \bS_i \cdot \bS_j \; \rgl \lgl \; \bS_n \cdot \bS_n \; \rgl \qquad
( i \neq n ,  ~  j \neq n  ) \; ,
$$
$$
\lgl \; \bS_j (\bS_j \cdot \bS_n) \bS_n \; \rgl =  
\lgl \; \bS_j \cdot \bS_j \; \rgl \lgl \; \bS_n \cdot \bS_n \; \rgl \qquad
( j \neq n  ) \;  .
$$
This leads to the matrix elements
$$
C_{ijmn} = M^4 \qquad ( i \neq n , ~ j \neq m , ~  j \neq n , ~ m \neq n ) \; ,
$$
$$
C_{jjmn} = S ( S + 1)  M^2 \qquad (  j \neq n , ~ j \neq m , ~ m \neq n ) \; ,
$$
$$
C_{ijnn} = S ( S + 1)  M^2 \qquad (   i \neq n , ~ j \neq n ) \; ,
$$
$$
C_{jjnn} = [\; S ( S + 1)\; ]^2  \qquad ( j \neq n ) \;   .
$$
Therefore we obtain
\be
\label{119}
 ||\; \hat C_2\; || = 2 [\; S ( S + 1 ) \; ]^2 + 4 ( N - 1 ) S ( S + 1 ) M^2 +
( N - 1 )^2 M^4 \;  .
\ee

The nonentangling counterpart of the second-order correlation matrix,
\be
\label{120}
 \hat C_2^\otimes = \hat C_1 \bigotimes \hat C_2 \;  ,
\ee
for which
\be
\label{121}
{\rm Tr} \hat C_2^\otimes = {\rm Tr} \hat C_2 = [\; S ( S + 1 ) N \; ]^2  ,
\ee
results in the norm
\be
\label{122}
||\; \hat C_2^\otimes \; || = ||\; \hat C_1\; ||^2 = 
 [\; S ( S + 1 ) \; ]^2 + 2 ( N - 1 ) S ( S + 1 ) M^2 + ( N - 1 )^2 M^4 \;  .
\ee

The norms are essentially different for the magnetically ordered state, when a nonzero 
magnetization is present, and for the paramagnetic state without average magnetization:
\begin{eqnarray}
\nonumber
||\; \hat C_1\; || = \left\{ \begin{array}{ll}
N M^2 , ~ & ~ | M | > 0 \\
S(S+1) , & ~  M = 0
\end{array} \; ,
\right.
\end{eqnarray}
\begin{eqnarray}
\nonumber
||\; \hat C_2\; || = \left\{ \begin{array}{ll}
N^2 M^4 , ~ & ~ | M | > 0 \\
2[\; S(S+1) \;]^2 , ~ & ~  M = 0
\end{array} \; ,
\right.
\end{eqnarray}
\begin{eqnarray}
\label{123}
||\; \hat C_2^\otimes\; || = \left\{ \begin{array}{ll}
N^2 M^4 , ~ & ~ | M | > 0 \\
\; [\; S(S+1) \;]^2 , ~ & ~  M = 0
\end{array} \; ,
\right.
\end{eqnarray}
For the order indices 
\be
\label{124}
\om(\hat C_n) = \frac{\log|\;\hat C_n\;|}{\log|\; {\rm Tr}\hat C_n \;|} \;   ,
\ee
we have
\begin{eqnarray}
\label{125}
\om(\hat C_1) = \om(\hat C_2) = 
\left\{ \begin{array}{ll}
1 , ~ & ~ | M | > 0 \\
0 , & ~  M = 0
\end{array} \; .
\right.
\end{eqnarray}
For the entanglement-production measure
\be
\label{126}
 \ep(\hat C_2) = \log\; \frac{||\; \hat C_2\;||}{||\;\hat C_2^\otimes\;||} \;  ,
\ee 
we find
\begin{eqnarray}
\label{127}
\ep(\hat C_2) = 
\left\{ \begin{array}{ll}
0 , ~ & ~ | M | > 0 \\
\log 2 , & ~  M = 0
\end{array} \; .
\right.
\end{eqnarray}
Here a large system, with $N \gg 1$ is assumed. 

Again we see that the entanglement production diminishes upon arising order and 
increases for a disordered state.

\section{Diagonal order}

In the previous Sections $10$ and $11$, the cases of phase transformations with the 
arising off-diagonal order are analyzed. There exists an opinion that the transitions 
with the arising diagonal order have to be treated differently, since the related reduced 
density matrices behave in a different way. However, in the approach based on the order 
indices, there is no difference in the method of treating any type of phase transition. 
What needed is to define the appropriate correlation matrix \cite{Yukalov_10}. In the 
present section, we illustrate this for the solid-liquid transition that is the most 
known transition exhibiting diagonal order. 

Under the solidification--melting phase transition, what changes is the particle density
\be
\label{128}
 \rho(\br) = \lgl \; \hat\rho(\br) \; \rgl  
\ee 
that is the statistical average of the density operator
\be
\label{129}
 \hat\rho(\br) \equiv \psi^\dgr(\br) \psi(\br) \;  .
\ee
In the liquid state, the particle density is constant in space, being equal to the 
average density
\be
\label{130}
 \rho = \frac{1}{V} \int \rho(\br) \; d\br = \frac{N}{V} \;  .
\ee
In the solid state, the density is nonuniform, having minima and maxima. 

We can label the points of the density maxima by ${\bf a}_j$, enumerating them by the 
index $j = 1,2,\ldots,N_L$, thus defining them by the condition
\be
\label{131}
 \max_\br \rho(\br) = \rho(\ba_j) \;  .
\ee
Generally, the points of maxima do not need to form a periodic structure, but can be 
randomly located, as in amorphous solids. For crystalline structures, the points of 
density maxima form a periodic crystalline lattice, where the set of $\{{\bf a}_j\}$
fixes lattice sites. Below, we shall consider a crystalline lattice, where the points 
of density maxima are identical to each other, so that 
\be
\label{132}
 \rho(\ba_j) = \rho(\ba) \;  ,
\ee
with ${\bf a}$ being any site from the set of lattice sites. This is not principal 
for the approach, but just simplifies some notations.      

Let us introduce the operator
\be
\label{133}
 \hat A(\ba_j) \equiv \hat\rho(\ba_j) - \rho \;  .
\ee
Its average is
$$
\lgl \;  \hat A(\ba_j) \; \rgl = \rho(\ba) - \rho \;   .
$$
With these operators, it is straightforward to define the correlation functions
\be
\label{134}
 D_{ij} = \lgl \;  \hat A(\ba_i) \hat A(\ba_j) \; \rgl \;  .
\ee
These functions play the role of matrix elements of the correlation matrix
\be
\label{135}
 \hat D_1 = [\; D_{ij} \; ] \;  ,
\ee
for which
\be
\label{136}
 D_{ij} = \lgl \;  \hat\rho(\ba_i) \hat\rho(\ba_j) \; \rgl - 
2\rho \rho(\ba) + \rho^2 \;  .
\ee

The trace of this matrix is
\be
\label{137}
 {\rm Tr} \hat D_1 = \sum_j D_{jj} = N_L \left[ \lgl \; \hat\rho(\ba)^2 \;\rgl
-  2\rho \rho(\ba) + \rho^2  \right] \; ,
\ee
with $N_L$ being the number of lattice sites. This trace can be rewritten as
\be
\label{138}
 {\rm Tr} \hat D_1 = N_L \left\{ [\; \rho(\ba) - \rho \; ]^2 + 
{\rm var}\hat\rho(\ba) \;  \right\} \;   ,
\ee
where 
$$
{\rm var}\hat\rho(\ba) \equiv \lgl \; \hat\rho(\ba)^2 \;\rgl - \rho(\ba)^2
$$
is the density variance at a lattice site. For different sites, one can use the Hartree
decoupling
$$
\lgl \; \hat\rho(\ba_i) \hat\rho(\ba_j) \; \rgl = 
\lgl \; \hat\rho(\ba_i) \; \rgl \lgl \; \hat\rho(\ba_j) \; \rgl 
\qquad 
( i \neq j)
$$
that is known to provide a good description for crystals 
\cite{Guyer_107,Zubov_108,Glyde_109,Yukalov_110,Zubov_111,Zubov_112}. Keeping in mind 
the lattice function in Equation (\ref{111}), for the norm of the matrix, we find
\be
\label{139}
||\; \hat D_1 \; || = \rho(\ba)^2 + {\rm var}\hat\rho(\ba) +
N_L [\; \rho(\ba) - \rho \; ]^2 \;  .
\ee

In this way, we come to the order index
$$
\om(\hat D_1) = \frac{\log||\;\hat D_1\;||}{\log|\;{\rm Tr}\hat D_1\;|} =
$$
\be
\label{140}
=
\frac{\log\{N_L[\rho(\ba)-\rho]^2+\rho(\ba)^2+{\rm var}\hat\rho(\ba)\} }
{\log N_L + \log\{[\rho(\ba)-\rho]^2+{\rm var}\hat\rho(\ba)\} } \;  .
\ee
For a solid, this reduces to
\be 
\label{141}
\om(\hat D_1) = 1 + \frac{2\log[\rho(\ba)-\rho]}{\log N_L} \; , 
\qquad
\rho(\ba) > \rho \;   ,
\ee
while for a liquid, it gives
\be
\label{142}
 \om(\hat D_1) = \frac{\log\{\rho(\ba)^2+{\rm var}\hat\rho(\ba)\}}{\log N_L} \; , 
\qquad
\rho(\ba) = \rho \; .
\ee
These expressions can be used for both, finite systems as well as for macroscopic 
systems with a large number of particles. Note that finite crystalline structures can 
be stable, or metastable, even in low dimensions, forming crystalline chains and planes 
\cite{Yukalov_113} for which the Lindemann stability criterion \cite{Lindemann_114} 
is valid. In the thermodynamic limit, we obtain
\begin{eqnarray}
\label{143}
 \om(\hat D_1) = \left\{ \begin{array}{ll}
1 , ~ & ~ {\rm solid} \\
0 , ~ & ~ {\rm liquid}
\end{array} 
\right. \qquad ( N_L \ra \infty ) \; .
\end{eqnarray}

Similarly, the characteristics of higher-order correlation matrices $\hat{D}_n$ can be 
calculated. The overall situation is analogous to the cases of phase transitions treated 
in Secs. $10$ and $11$.

\section{Dynamical effects}

Order indices and entanglement production can vary with time. Time dependence can come 
through equations of motion. If the operator obeys a unitary evolution, the order 
indices do not change with time, although the entanglement production can vary since 
the non-entangling operator, generally, does not follow the unitary evolution. Reduced 
density matrices and correlation matrices, in general, also are not governed by unitary 
time dependence. Therefore the order indices and entanglement-production measure, say 
for reduced density matrices of nonequilibrium systems, do depend on time,
$$
\om(\hat\rho_n(t) ) = 
\frac{\log||\;\hat\rho_n(t)\;||}{\log|\;{\rm Tr}\hat\rho_n(t)\;|} \; ,
$$
\be
\label{144}
\ep(\hat\rho_n(t) ) = 
\log\; \frac{||\;\hat\rho_n(t)\;||}{||\;\hat\rho_n^\otimes(t)\;||} \; .
\ee

\subsection{Multitrap multimode states}

As an example, let us consider the case of multiple traps filled by Bose--Einstein 
condensate. This can be realized with an optical lattice having deep wells, where 
condensate clouds are located. Thus, cold rubidium $^{87}$Rb atoms were loaded 
\cite{Hadzibabic_125,Cennini_115} into an optical lattice, with adjacent sites 
spaced so that these sites were practically independent, with the tunneling time 
between sites above $10^{18}$ s. The number of lattice sites was typically between 
$5$ to $35$. The number of condensed atoms in each site could be varied between 
about $200$ to $10^4$. Shaking the lattice as a whole, it is possible to create in 
each lattice site multimode states of excited Bose--Einstein condensates 
\cite{Yukalov_93,Yukalov_94,Yukalov_95,Yukalov_96,Kivshar_116,Yukalov_117,Proukakis_118,
Yukalov_119,Liu_120}. Such states are described by the statistical operator
\be
\label{145}
 \hat\rho_n(t) = \sum_{k=1}^{N_M} n_k(t)\; | \; k k \ldots k \; \rgl
\lgl \;  k k \ldots k \; | \; ,
\ee
where $n_k$ is the fraction of atoms in the $k$-th mode, $N_M$ is the number of modes 
in a lattice site, whose number is $N_L$. The normalization of the statistical operator 
to one, requires the validity of the summation
$$
\sum_{k=1}^{N_M} n_k = 1 \; .
$$
The entanglement-production is quantified by the measure
\be
\label{146}
\ep(\hat\rho(t) ) = ( 1 - N_L ) \log \;\sup_k \; n_k(t)
\ee
varying in the range
\be
\label{147}
 0 \leq \ep(\hat\rho(t) ) \leq (N_L - 1) \log N_M \;  .
\ee
The temporal evolution of the measure was studied for two and three modes
\cite{Yukalov_121,Yukalov_122,Yukalov_123,Yukalov_124}.   
  
For the realization of the multitrap multimode states it is important that the traps 
be identical, so that the transition frequencies between the ground-state condensate 
and the generated coherent mode be the same in all traps. Only then it is feasible 
to shake the lattice with a frequency that would be in resonance with the transition 
frequency of all traps, which is necessary for the generation of the same mode in 
these traps. A collection of different traps, such as effective potential wells in 
a random matter, where the so-called Bose glass can arise 
\cite{Fisher_126,Palencia_127,Konenberg_128} is not appropriate, since these effective 
wells are not identical, hence are characterised by different transition frequencies.

\subsection{Entangling by evolution operators}

Entanglement production by the evolution operator 
\be
\label{148}
\hat U(t) = e^{-iH t}
\ee
was studied in Refs. \cite{Yukalov_44,Yukalov_129}. The operator norm was defined as 
the Hilbert--Schmidt norm. Calculations were performed for finite-site Heisenberg and 
Ising Hamiltonians. Depending on the system parameters, the evolution of the 
entanglement-production measure 
\be
\label{149}
\ep(\hat U(t)) = \log\; \frac{||\;\hat U(t)\;||}{||\;\hat U^\otimes(t)\;||}
\ee
is periodic or quasiperiodic.

\section{Coherence phenomena}

Coherence phenomena occurring in nonequilibrium systems can be treated as a kind 
of dynamic phase transitions. Therefore the temporal behavior of these coherent 
phenomena can also be accompanied by drastic changes in the order indices and 
entanglement production. Examples of these phenomena are given by the dynamics of 
strongly nonequilibrium spin systems \cite{Yukalov_139,Yukalov_140} and radiating 
systems \cite{Yukalov_130,Yukalov_131,Yukalov_132}. Being many-particle collections, 
these systems allow for considering order indices and entanglement production 
characterizing their temporal behavior. To describe nonequilibrium processes in 
these systems, it is convenient to resort to quasispin representation 
\cite{Yukalov_130,Yukalov_131,Yukalov_132} based on the quasispin operators
$$
S_j^z = \frac{1}{2}\; \sgm_j^z \; , \qquad 
S_j^\pm = \frac{1}{2}\; \sgm_j^\pm \;  ,
$$
where the index $j$ enumerates particles, or spins, and $\sgm_j^\al$ are Pauli 
matrices. The main observable quantities are the fractional population imbalance
(or longitudinal spin polarization)
\be
\label{150}
s \equiv \frac{2}{N} \sum_{j=1}^N \lgl \; S_j^z \; \rgl
\ee
and the dimensionless coherence intensity or radiation intensity
\be
\label{151}
  w \equiv \frac{4}{N^2} \sum_{i\neq j}^N \lgl \; S_i^+ S_j^- \; \rgl 
\ee
that are functions of time coming from the equations of motion. Summation is over 
all atoms or spins. It is useful to mention that in nonequilibrium processes, such 
as radiation, the population imbalance $s$ is never identically equals minus one, 
which would mean an equilibrium state, when the coherence intensity $w$ would be 
zero.   

Since there are two physically different operators, one related to coherence 
intensity and the other describing the atomic population imbalance (spin polarization), 
it is convenient to consider two different types of correlation functions 
\cite{Yukalov_133}. Dealing with the population-imbalance operators requires to 
define correlation functions using the $z$-spin operators. Thus, it is possible to 
introduce the first-order correlation operator
\be
\label{152}
 \hat Q_1 = [\; Q_{ij} \; ] \; ,
\ee
with the matrix elements
\be
\label{153}
 Q_{ij} \equiv \lgl \; S_i^z S_j^z \; \rgl \;  .
\ee
The second-order correlation operator is
\be
\label{154}
\hat Q_2 = [\; Q_{ijmn} \; ] \;  ,
\ee
having the matrix elements
\be
\label{155}
Q_{ijmn} \equiv \lgl \; S_i^z S_j^z S_m^z S_n^z \; \rgl \; .
\ee

Similarly, it is straightforward to define the correlation operators composed of the 
ladder spin operators, the first order 
\be
\label{156}
\hat R_1 = [\; R_{ij} \; ] \;  ,
\ee
with the elements
\be
\label{157}
R_{ij} \equiv \lgl \; S_i^+ S_j^- \; \rgl \;   ,
\ee
and the second order
\be
\label{158}
\hat R_2 = [\; R_{ijmn} \; ] \;   ,
\ee
with the matrix elements
\be
\label{159}
R_{ijmn} \equiv \lgl \; S_i^+ S_j^+ S_m^- S_n^- \; \rgl \;   .
\ee
To calculate the traces and norms of the correlation operators, we shall need the 
following equalities for coinciding locations, 
$$
S_j^+ S_j^+ =  S_j^- S_j^- = 0 \; \qquad 
S_j^z S_j^z = \frac{1}{4} \; ,
$$
$$
S_j^+ S_j^- =  \frac{1}{2} + S_j^z \; , \qquad 
S_j^- S_j^+ = \frac{1}{2} - S_j^z \; ,
$$
$$
S_j^+ S_j^z =  -\; \frac{1}{2} \; S_j^+ \; , \qquad 
S_j^- S_j^z = \frac{1}{2}\; S_j^- \; ,
$$
$$
S_j^z S_j^+ =  \frac{1}{2} \; S_j^+ \; , \qquad 
S_j^z S_j^- = -\; \frac{1}{2}\; S_j^- \; .
$$

Let us start with the operators $\hat{Q}_n$. For the traces, we have
\be
\label{160}
 {\rm Tr}\hat Q_1 = \sum_{j=1}^N Q_{jj} = \frac{N}{4} \; , 
\qquad 
 {\rm Tr}\hat Q_2 = \sum_{jn} Q_{jnnj} = \frac{N^2}{16} \; .
\ee
The partially traced operators
\be
\label{161}
\hat Q_1^{(\al)} = {\rm Tr}_{\cH/\cH_\al}\hat Q_2 = \frac{N}{4}\; \hat Q_1 
\qquad
(\al = 1,2 ) \; \;
\ee
give the related nonentangling factor operator
\be
\label{162}
 \hat Q_1^\otimes = \frac{Q_1^{(1)}\bigotimes Q_1^{(2)}}{{\rm Tr}\hat Q_2} = 
\hat Q_1 \bigotimes \hat Q_2 \;  .
\ee
For the norms, we find
$$
||\; \hat Q_1\; || = \frac{1}{4} \; \left( 1 + N s^2 \right) \; , 
$$
$$
||\; \hat Q_2\; || = \frac{1}{8} + \frac{N}{4}\; s^2 + \frac{N^2}{16}\; s^4 \; , 
$$
\be
\label{163}
||\; \hat Q_2^\otimes \; || = ||\; \hat Q_1\; ||^2 = 
\frac{1}{16}\; \left( 1 + N s^2\right)^2 \; .
\ee

Calculating the order indices and entanglement-production measure, we simplify the 
final expressions by keeping in mind a large number of atoms $N \gg 1$. Then we get 
the order indices
\begin{eqnarray}
\label{164}
\om(\hat Q_1) = \om(\hat Q_2) =\left\{ \begin{array}{ll}
1 , ~ & ~ s \neq 0 \\
0 , ~ & ~ s = 0
\end{array} \right.
\end{eqnarray}
and the entanglement-production measure
\begin{eqnarray}
\label{165}
\ep(\hat Q_2) = \left\{ \begin{array}{ll}
0 , ~ & ~ s \neq 0 \\
\log 2 , ~ & ~ s = 0
\end{array} 
\right. \; .
\end{eqnarray}
The behavior of the order indices here is similar to the case of magnetic systems. 
However the physical nature of the arising order is rather different from the latter. 
In the process of coherent dynamics or radiation, the ordering occurs for a short time, 
since $s=s(t)$ is a function of time. Here it is an example of a dynamical order, 
while in a magnet, it is a stationary order. Similarly to the stationary magnetic 
order, the dynamic order 
 also reduces the measure of entanglement production.      

For the correlation operators related to the coherence intensity, we have the traces
\be
\label{166}
{\rm Tr}\hat R_1 = \sum_j \lgl \; S_j^+ S_j^- \; \rgl = \frac{N}{2} \; (1 + s) \; ,
\qquad
{\rm Tr}\hat R_2 =  \frac{N^2}{4} \; (1 + s)^2 \; .
\ee
The partially traced operators
\be
\label{167}
 \hat R_1^{(\al)} = \frac{N}{2} \; ( 1 + s ) \hat R_1 \qquad ( \al = 1,2 )   
\ee
result in the nonentangling factor operator
\be
\label{168}
\hat R_2^\otimes = \frac{\hat R_1^{(1)}\bigotimes \hat R_1^{(2)}}{{\rm Tr}\hat R_2} 
= \hat R_1 \bigotimes \hat R_1  \; .
\ee
We find the following norms
$$
||\;  \hat R_1 \; || = \frac{1}{2}\; (1 + s) + \frac{N}{4}\; w \; , \qquad
||\;  \hat R_1^{(\al)} \; || = \frac{N}{2}\; (1 + s) ||\;  \hat R_1 \; ||  \; ,
$$
\be
\label{169}
||\;  \hat R_2^\otimes \; || = ||\;  \hat R_1 \; ||^2 \; , 
\qquad
||\;  \hat R_2 \; || = \frac{1}{2}\; (1 + s)^2 + \frac{N}{2}\; (1 + s) w +
\frac{N^2}{16}\; w^2 \; .
\ee

Thus, for large $N \gg 1$, we obtain the order indices
\begin{eqnarray}
\label{170}
\om(\hat R_1) = \om(\hat R_2) = \left\{ \begin{array}{ll}
1 , ~ & ~ w \neq 0 \\
0 , ~ & ~ w = 0 
\end{array} \right.
\end{eqnarray}
and the entanglement-production measure
\begin{eqnarray}
\label{171}
\ep(\hat R_2) = \left\{ \begin{array}{ll}
0 , ~ & ~ w \neq 0 \\
\log 2 , ~ & ~ w = 0 
\end{array} \right. \; .
\end{eqnarray}

These quantities essentially depend on the coherence intensity $w = w(t)$ that plays 
here the role of a dynamic order characteristic. A nonzero $w$ signifies the appearance 
of a coherent dynamic order. Sometimes, the occurrence of the coherent dynamic order is 
paralleled with the nonequilibrium magnon condensation 
\cite{Volovik_134,Bunkov_135,Bunkov_136}. Note that in equilibrium systems magnons 
cannot condense \cite{Yukalov_137,Birman_138}.

\section{Conclusion}

We have surveyed the meaning and applications of two concepts, order indices and 
entanglement production. The concept of order indices generalizes that of order 
parameters. The latter usually describe long-range order and are well defined only 
for macroscopic systems in the thermodynamic limit. However the order indices can be 
defined for any system, whether in the thermodynamic limit or finite. They characterize 
all types of order, long-range, mid-range, or short-range, making it possible not 
merely distinguishing the qualitative types of orders, but prescribing a measure 
uniquely quantifying the level of ordering and being applicable to equilibrium as 
well as to nonequilibrium systems.    

The entanglement production describes how much entanglement is produced by an 
operator. This should be distinguished from the notion of entanglement describing 
the operator structure. It is possible to say that entanglement, characterizing the 
operator structure, is a static notion, while the entanglement production, describing 
the operator action, is a kind of a dynamic notion. The entanglement production can 
be quantified by a measure that is valid for any system, bipartite or multipartite, 
equilibrium or not.  

It turns out that order indices and the entanglement-production measure are closely 
related with each other. As a rule, the larger the level of order in a system, the 
smaller the entanglement-production measure. The use of these concepts helps to better 
understand the properties of the studied systems and to more efficiently employ them 
in applications.

In the review, we considered the cases that could be treated analytically. Dealing 
with finite systems, it is usually necessary to resort to numerical calculations. 
Thus powerful numerical methods have been developed for studying the dynamics of 
finite Bose systems \cite{Alon_141,Alon_142,Lode_143}. Hopefully, the described 
notions could be successfully employed for the analysis of finite systems by applying 
numerical methods.     

\section*{Acknowledgments}

The author is grateful for many useful discussions to E.P. Yukalova.

\end{document}